\RequirePackage{fix-cm}
\documentclass[twocolumn,epjc3]{svjour3}  
\smartqed  
\RequirePackage{graphicx}
\RequirePackage{comment}
\usepackage{color}
\usepackage{lineno}
\usepackage{url}
\usepackage{amsmath}
\usepackage{breqn}
\usepackage{amsfonts}

\newcommand{\LMO}{Li$_2$MoO$_4$}
\newcommand{\LMOenr}{Li$_2$$^{100}$MoO$_4$}
\newcommand{\DBD}{$0\nu$DBD}
\newcommand{\TDBD}{$2\nu$DBD}

\journalname{Eur. Phys. J. C}

\begin{document}

\title{Phonon and light read out of a {\LMO} crystal with multiplexed kinetic inductance detectors.}

\author{N.~Casali\thanksref{e1,INFNRoma}
\and  L.~Cardani\thanksref{INFNRoma}
\and I.~Colantoni\thanksref{CNRRoma,INFNRoma}
\and A.~Cruciani\thanksref{INFNRoma}
\and S.~Di~Domizio\thanksref{Genova,INFNGenova}
\and M. Martinez\thanksref{Spagna}
\and G.~Pettinari\thanksref{IFN}
\and M.~Vignati\thanksref{INFNRoma}}

\institute{
INFN - Sezione di Roma, Roma I-00185 - Italy\label{INFNRoma}
\and
Consiglio Nazionale delle Ricerche, Istituto di Nanotecnologia  (CNR - NANOTEC), c/o Dip. Fisica, Sapienza Universit\`{a} di Roma, Roma I-00185 - Italy\label{CNRRoma}
\and
Dipartimento di Fisica, Universit\`{a} di Genova, Genova I-16146 - Italy\label{Genova}
\and
INFN - Sezione di Genova, Genova I-16146 - Italy\label{INFNGenova}
\and
Laboratorio de F\'isica Nuclear y Astropart\'iculas, Universidad de Zaragoza, C/ Pedro Cerbuna 12, 50009 and Fundaci\'on ARAID, Av. de Ranillas 1D, 50018 Zaragoza, Spain\label{Spagna}
\and
Consiglio Nazionale delle Ricerche, Istituto di Fotonica e Nanotecnologie (CNR - IFN), Via Cineto Romano 42, 00156, Roma - Italy\label{IFN}
}

\thankstext{e1}{e-mail: nicola.casali@roma1.infn.it}


\date{Received: date / Accepted: date}

\maketitle
\begin{abstract}
Molybdenum based crystals such as {\LMO} and CaMoO$_4$ are emerging as leading candidates for next generation experiments searching for neutrino-less double beta decay with cryogenic calorimeters (CUPID, AMoRE). The exquisite energy resolution and high radio-purity of these crystals come at the cost of a potentially detrimental background source: the two neutrinos double beta decay of $^{100}$Mo. Indeed, the fast half-life of this decay mode, combined with the slow response of cryogenic calorimeters, would result in pile-up events in the energy region of interest for neutrino-less double beta decay, reducing the experimental sensitivity. This background can be suppressed using fast and high sensitivity cryogenic light detectors, provided that the scintillation time constant itself does not limit the time resolution. We developed a new detection technique exploiting the high sensitivity, the fast time response and the multiplexing capability of Kinetic Inductance Detectors. We applied the proposed technique to a $2\times2\times2$~cm$^3$ {\LMO} crystal, which was chosen as baseline option for CUPID. We measured simultaneously both the phonon and scintillation signals with KIDs. We derived the scintillation time constant of this compound at millikelvin temperatures obtaining $\tau_{scint} = 84.5\pm4.5\rm{(syst)}\pm1.0\rm{(stat)}$~$\mu$s, constant between 10 and 190~mK.

\end{abstract}

\section{Introduction}
\label{intro}
Neutrinos are the particles of the Standard Model that best hide their fundamentals properties. Despite being the most abundant particles in the Universe, their low mass and cross section make them also the most elusive ones. The mass scale as well as their nature (Dirac or Majorana) are still unknown, more than 60 years after their first detection. Demonstrating the existence of the neutrino-less double beta decay ({\DBD})~\cite{Furry} would offer the possibility to answer both questions~\cite{Feruglio:2002af,Strumia:2005tc}. This reaction is a hypothesized nuclear transition in which a nucleus decays emitting two electrons and no neutrinos: (A,Z)$\rightarrow$(A,Z+2)+2e$^-$. The signal produced would consist of two electrons with a total kinetic energy equal to the Q-value of the transition. The current lower limits on the half-life of this process are of the order of $10^{24}-10^{26}$~yr, depending on the nucleus. On the contrary, the two neutrinos double beta decay ({\TDBD}), is a well establish second order process of the Standard Model: (A,Z)$\rightarrow$(A,Z+2)+2e$^-$+2$\bar{\nu}$. The signal consists of two electrons with a continuous energy spectrum up to the Q-value of the transition. This process was measured for 11 nuclei with half-lives ranging from $10^{18}-10^{24}$~yr~\cite{Barabash:2019nnr}.

Next generation experiments searching for {\DBD} are conceived to improve their sensitivity by increasing the source mass and reducing the background. The CUPID (CUORE Upgrade with Particle Identification) interest group~\cite{Wang:2015taa} proposes a next generation {\DBD} experiment upgrading the cryogenic calorimetric technique developed by the CUORE experiment~\cite{Artusa:2014lgv}, with the aim of decreasing the background from $10^{-2}$ to $10^{-4}$~counts/(keV~kg~yr). The main background of CUORE comes from $\alpha$ particles, as described in Ref.~\cite{Alduino:2017qet}. Therefore, the first step to achieve the goal of CUPID consists in rejecting such kind of interactions. This discrimination is achieved by coupling each calorimeter to a light detector in order to disentangle $\alpha$ from $\beta/\gamma$ interactions, thanks to the different light yield and time development of scintillation light. The potential of the dual read out was convincingly proved by the pilot experiment CUPID-0~\cite{Azzolini:2018tum}, with an array of ZnSe scintillating calorimeters 95\% enriched in $^{82}$Se, and by the LUMINEU~\cite{Armengaud:2017hit} and AMoRE~\cite{Alenkov:2019jis} projects, with similar detectors based respectively on {\LMO} 95\% enriched in $^{100}$Mo and CaMoO$_4$ depleted in $^{48}$Ca and enriched in $^{100}$Mo.
{\LMO} offers higher radio-purity and better energy resolution compared to ZnSe and, in contrast to CaMoO$_4$ does not require a further depletion of Ca for applications in double beta decay searches. For these reasons, it was chosen as baseline for the CUPID project. Unfortunately, $^{100}$Mo, with a half-life of ($7.1\pm0.4$)$\times10^{18}$~y~\cite{Barabash:2011} is one of the fastest {\TDBD} emitter: a {\LMO} enriched crystal has an intrinsic activity of about 10~mBq/kg coming from the $^{100}$Mo {\TDBD} decay. The typical time response of a cryogenic calorimeter exploiting the NTD Ge thermistors technology is 10~-~100~ms for the rise time and 50~-~400~ms for the decay time~\cite{Artusa:2014lgv,Azzolini:2018yye}, leading to a non-negligible probability of pile-up of two {\TDBD} events. The estimated background produced by {\TDBD} pile-up in the {\DBD} energy region by a {\LMOenr} enriched crystal of 300~g is of the order of $1.1\times10^{-4}$ counts/(keV~kg~yr)~\cite{Chernyak1}. This background matches the requirements of CUPID but it would prevent a further sensitivity enhancement.

While an R\&D devoted to the size of the NTD Ge thermistor could be beneficial, it is hard to envision that it would largely improve the time resolution. Nevertheless, two events that are indistinguishable by the NTD Ge of the {\LMO} crystal could be disentangled studying its scintillation signal. Indeed, since the time constants of cryogenic calorimeters decrease abruptly with the mass of the absorber, we can use the faster response of the light detector (50 times less massive that {\LMO} crystal) to separate pile-up events. For example, if we consider the cryogenic light detectors used by the CUPID-0 experiment (4.4 cm in diameter and 170~$\mu$m thick Ge slab with a weight of about 6~grams, read by NTD Ge thermistors) rise and decay times result in 4 and 8 ms, respectively~\cite{Azzolini:2018yye}. For this reason, the current studies for the suppression of the {\TDBD} random coincidences pile-up are focused on the scintillation light measured by the light detector~\cite{Chernyak2}. To our knowledge, the fastest light detectors based on NTDs show a time response of 0.5-1 ms~\cite{Barucci:2019ghi}. A further improvement can be attained, but it would make sense only if the scintillation constant of the {\LMO} itself does not limit the time resolution. Nowadays the scintillation time constant of {\LMO} below 20~K is not known~\cite{CHEN2018225}. A slow time constant (1~ms) would limit the time resolution, preventing a further background suppression. 

In this contribution, we propose a new technological approach for the dual read out of light and phonon in a scintillating cryogenic calorimeter exploiting the Kinetic Inductance Detector (KID) technology. The fast time response and high sensitivity of KIDs allow us to measure for the first time the scintillation time constant at 10 mK of a {\LMO} crystal.


\section{Phonon-Mediated Kinetic Inductance Detectors}
\label{sec:KID}
In the last years a new technology, first developed for astrophysical applications, was proposed for cryogenic light detection: the Kinetic Inductance Detectors~\cite{bay}. In a superconductor cooled well below the critical temperature, the resistance to DC current is zero because the electrons, bounded in Cooper pairs, are not scattered by the lattice. If an AC current is applied, the superconductor shows the so called Kinetic Inductance, which is created by the inertia of the Cooper pairs that take some time (because of their mass) to change direction following the one of the applied EM field. Coupling a capacitance $C$ to the superconductor, a resonator with a high quality factor ($Q\sim10^5$) can be realized. The resonant frequency is given by $f_0=1/(2\pi\sqrt{LC})$, where $L$ depends on the density of Cooper pairs. Since their binding energy is of the order of hundreds of $\mu$eV, even a small energy release in the superconductor is able to break a high number of Cooper pairs, resulting in a change of the kinetic inductance $L$. The signal is obtained by exciting the circuit at the resonant frequency, and by monitoring the phase and amplitude variations, induced by energy releases, of the wave transmitted past the resonator. 
These devices were successfully applied in astro-physics providing outstanding performances in terms of energy resolution (eV), fast time response (10~$\mu$s), and multiplexing capability: many KIDs can be coupled to the same read out line (up to 1000 with 4~kHz bandwidth per detector~\cite{vanRantwijk:2015sta}) , and be simultaneously monitored making them resonate at slightly different frequencies by adjusting the layout of the capacitor and/or inductor of the circuit~\cite{Mazin:2010pz}.

The only limit for application in particle physics is their poor active surface, which is of few mm$^2$. This limitation can be overcome exploiting the phonon-mediated approach proposed by Swenson et al.~\cite{Swenson:2010} and Moore et al.~\cite{Moore:2012}: KIDs are deposited on an insulating substrate featuring a surface of several cm$^2$. Interactions in the substrate produce phonons, which travel in the substrate itself until a fraction of them are absorbed by the KID.

\begin{figure}[!t]
\begin{centering}
\includegraphics[width=8.5cm]{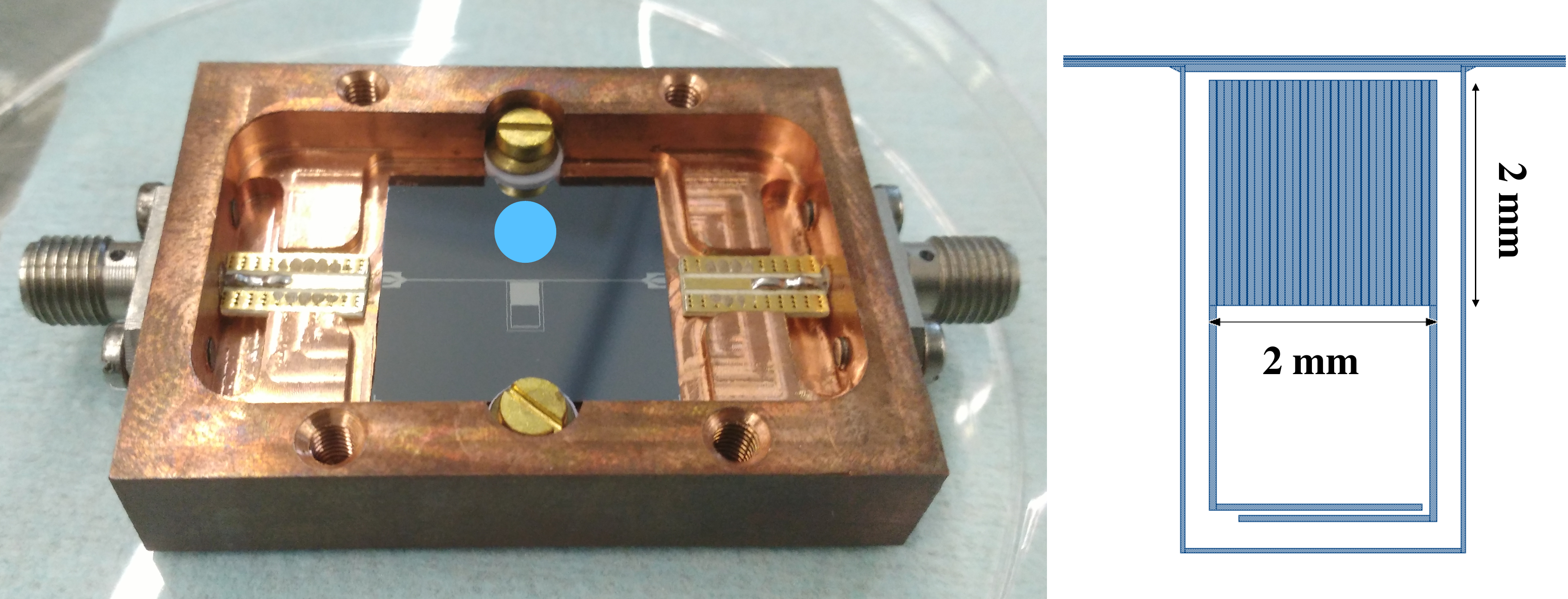}
\caption{Left: The light detector, consisting of a KID on a $2\times2$~cm$^2$ silicon substrate, is assembled in a copper structure by 2 PTFE supports. The circle shows the position and size of the spot of the optical system used to illuminate and calibrate the detector. Right: The KID inductor (30 strips of 62.5~$\mu$m$\times2$~mm size) features an active area of 3.75~mm$^2$.}
\label{fig:LD}
\end{centering}
\end{figure}

In this work, we apply the phonon mediated technique to a $2\times2\times2$~cm$^3$ {\LMO} scintillating calorimeter for the dual read out of phonons and scintillation light signals. Concerning the phonons read out, we propose for the first time to use KIDs coupled to a macro-calorimeter. Concerning the light measurement, we took advantage from the experience of the CALDER (Cryogenic wide Area Light Detector with Excellent Resolution) project~\cite{Battistelli:2015vha}, which successfully developed cryogenic light detector based on the KID technology. In the first phase of the R\&D activities, a Si substrate ($2\times2$~cm$^2$ and 300~$\mu$m thick) was used as phonon mediator; the superconductor resonator was made by aluminum, a well-known material for KID application. 
The best prototype developed in this first phase (shown in Fig.~\ref{fig:LD}) consists of one aluminum resonator 60~nm thick as this thickness value offers a good compromise between detector sensitivity and quality of the superconductor. The detector reached a baseline energy resolution of about 80~eV~RMS, constant in a temperature range from 10~mK up to 200~mK~\cite{Bellini:2016lgg}. Because of its robust performance we chose this prototype, even if CALDER proved that more sensitive superconductors provided an even better energy resolution~\cite{cruciani2018}.

\section{Detector set up}
\begin{figure}[!t]
\begin{centering}
\includegraphics[width=8.5cm]{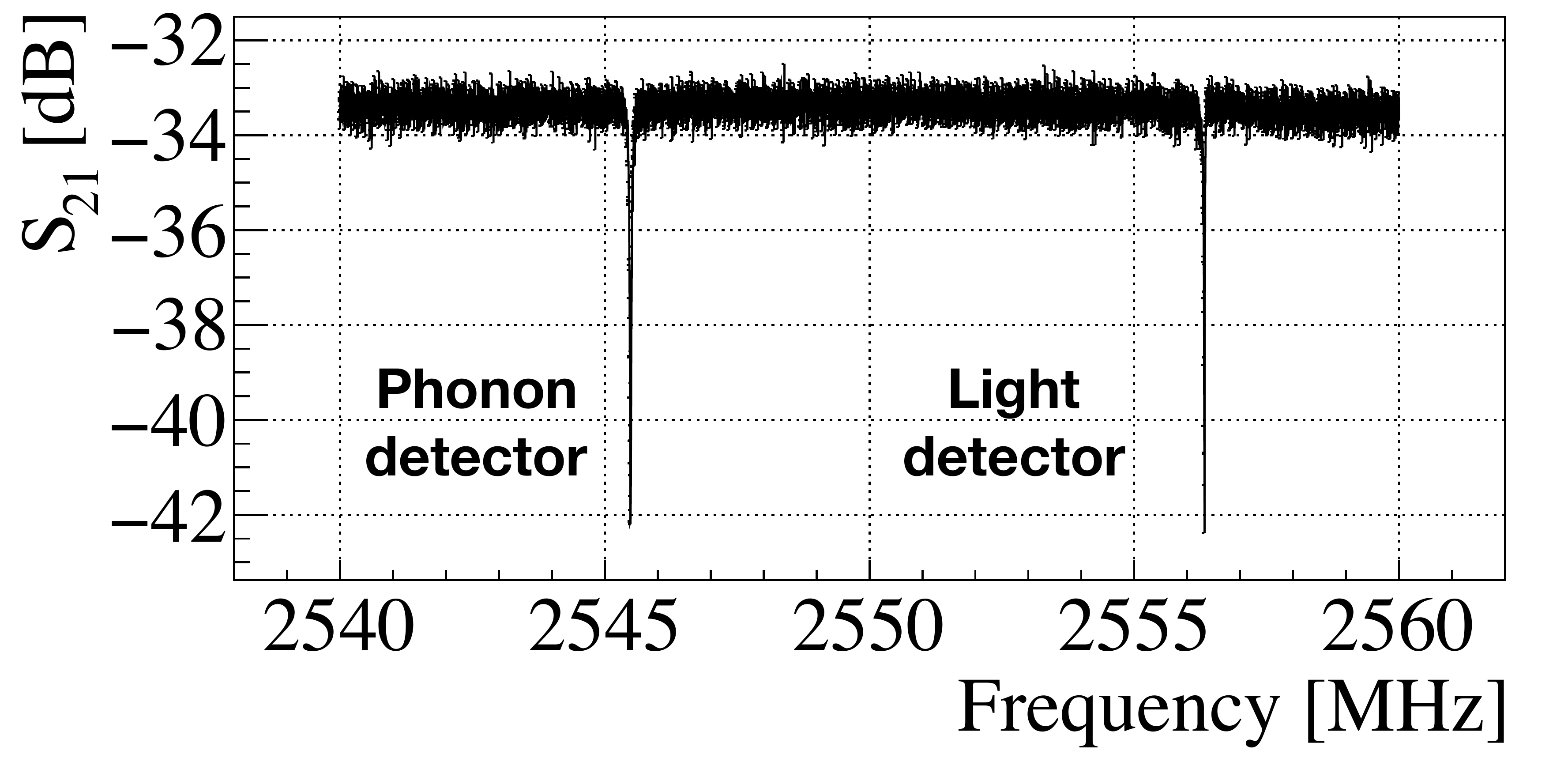}
\caption{Transmitted signal (S$_{21}$) as results from a frequency sweep of the electronic line where the two KIDs were coupled. The two resonances, corresponding respectively to the phonon and light detector of the {\LMO} crystal, can be easily recognized.}
\label{fig:line}
\end{centering}
\end{figure}

We face to a $2\times2\times2$~cm$^3$ {\LMO} crystal of 24.2~grams the KID based light detector showed in Fig.~\ref{fig:LD}. 
In order to select only scintillation light events coming from {\LMO} and disentangle them from direct energy deposition coming from environmental radioactivity ($\alpha$, $\beta$, $\gamma$ and $\mu$), we measured also the phonons produced in the {\LMO} crystal. To efficiently select coincidence events between the light detector and the {\LMO} crystal, the time resolutions of the two detectors must be comparable. For this reason, we instrumented also the {\LMO} crystal with a second KID.
\begin{figure}[!t]
\begin{centering}
\includegraphics[width=8.5cm]{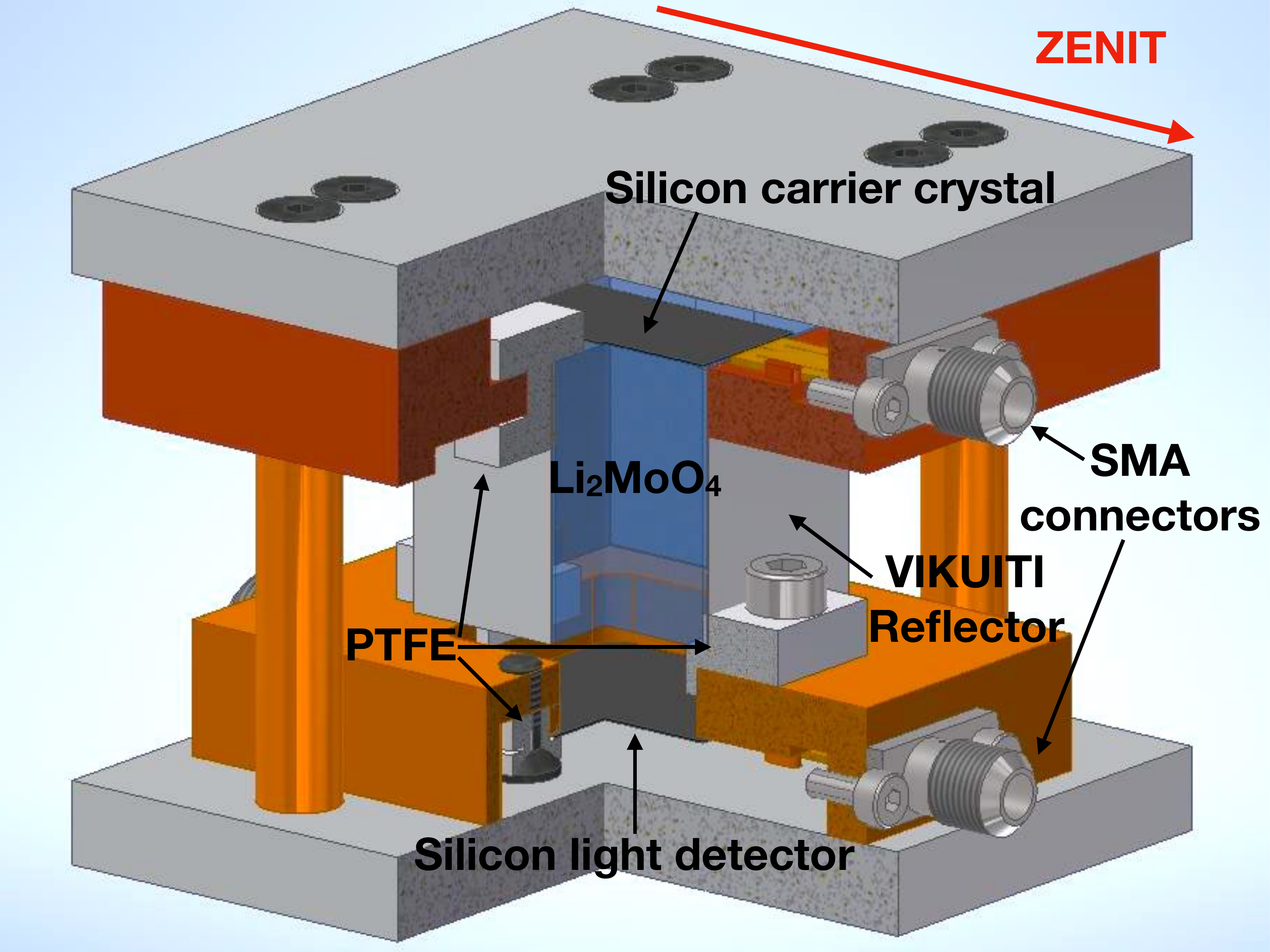}
\caption{The {\LMO} crystal is held by means of four PTFE supports and surrounded by ViKUITI reflective foils in order to increase the light collection efficiency. On one face of the crystal we glued the $2\times1$~cm$^2$ 300~$\mu$m thick silicon carrier crystal; the opposite face was monitored by the light detector shown in Fig.~\ref{fig:LD}. Both the resonators of the light and phonon channel have been realized at CNR - IFN in Rome. The detectors are oriented according to the red arrow visible on the top of the pictures. This orientation is the one that minimizes the number of vertical CRs crossing the Si substrates. The optical fiber that drives the LED pulses on the light detector is located on the other side with respect to the {\LMO} crystal.}
\label{fig:Setup}
\end{centering}
\end{figure}
Because of its hygroscopicity and other technical difficulties in depositing thin metallic layers on the {\LMO} crystal, we used the carrier approach: an Al KID, identical to the resonator depicted in Fig.~\ref{fig:LD} (except for one capacitor finger that was reduced to tune the resonance frequency) was evaporated on a Si substrate, $2\times1$~cm$^2$ 300~$\mu$m thick, which was then glued with EPO-TEK 301-2 on the {\LMO} crystal surface. The Si substrate works as carrier, transporting the phonons produced in the {\LMO} crystal by interacting particles to the KID deposited on it. This is the same technique that the CRESST experiment~\cite{cresst} successfully applied to transition edge sensors~\cite{tes}. The layout of the experimental set up is showed in Fig~\ref{fig:Setup}. Thank to the multiplexing capability of KIDs we coupled to the same read-out line both the phonon and light detectors. The signal transmitted past the two resonators is showed in Fig.~\ref{fig:line}. This is the first time in which the carrier approach is applied to KIDs, and in which a macro-calorimeter is operated with KIDs. We also faced to the light detector an optical fiber, in order to illuminate with a 400~nm room temperature LED lamp the light detector (see Fig~\ref{fig:LD}). We exploited such light pulses to energy calibrate the light detector and also study the KID response function to fast burst (100~ns) of optical photons.
The devices were mounted in the coldest point of a dry $^3$He/$^4$He dilution refrigerator\footnote{Oxford Instruments, Dry Dilution Refrigerator Triton 400.} and cooled down to 10~mK. The resonators were coupled in series to the read out line, excited and probed with a combination of two monochromatic tones, each one tuned at the resonant frequencies $f^1_0=2.5455$~GHz ({\LMO} crystal) and $f^2_0=2.5563$~GHz (light detector) (see Fig.~\ref{fig:line}). The output signals were fed into a CITLF3 SiGe cryogenic low noise amplifier~\cite{ampli}, down-converted at room temperature using a superheterodyne electronics and then digitized with an acquisition card at a sampling frequency of 500 kSPS~\cite{Bourrion:2011gi}. 
\begin{figure}[!b]
\begin{centering}
\includegraphics[width=8.5cm]{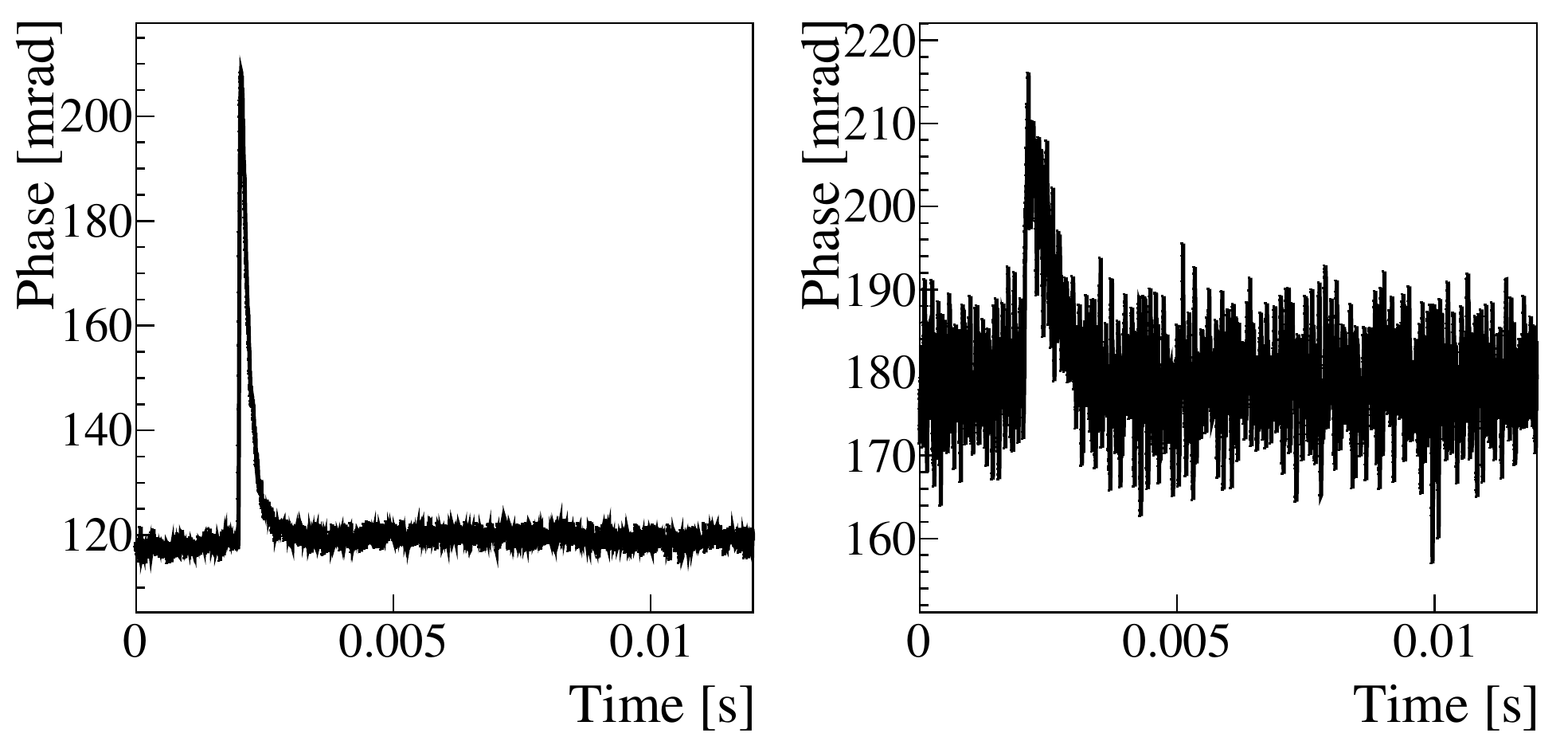}
\caption{Typical coincidence pulses triggered on phonon channel (left) and light channel (right). Because of the better signal-to-noise ratio of the phonon channel we used it as trigger for the light detector.}
\label{fig:Pulsi}
\end{centering}
\end{figure}
In Fig.~\ref{fig:Pulsi} the typical phase pulses filtered with a 180~kHz low pass filter are shown. Because of the better signal-to-noise ratio of the phonon channel, we used it to trigger both detectors. We used the maximum of the pulse as estimator of the energy release. To better evaluate this parameter we applied a software matched filter algorithm~\cite{Gatti:1986cw,Radeka:1966} to the acquired pulses.
\begin{figure}[!t]
\begin{centering}
\includegraphics[width=8.5cm]{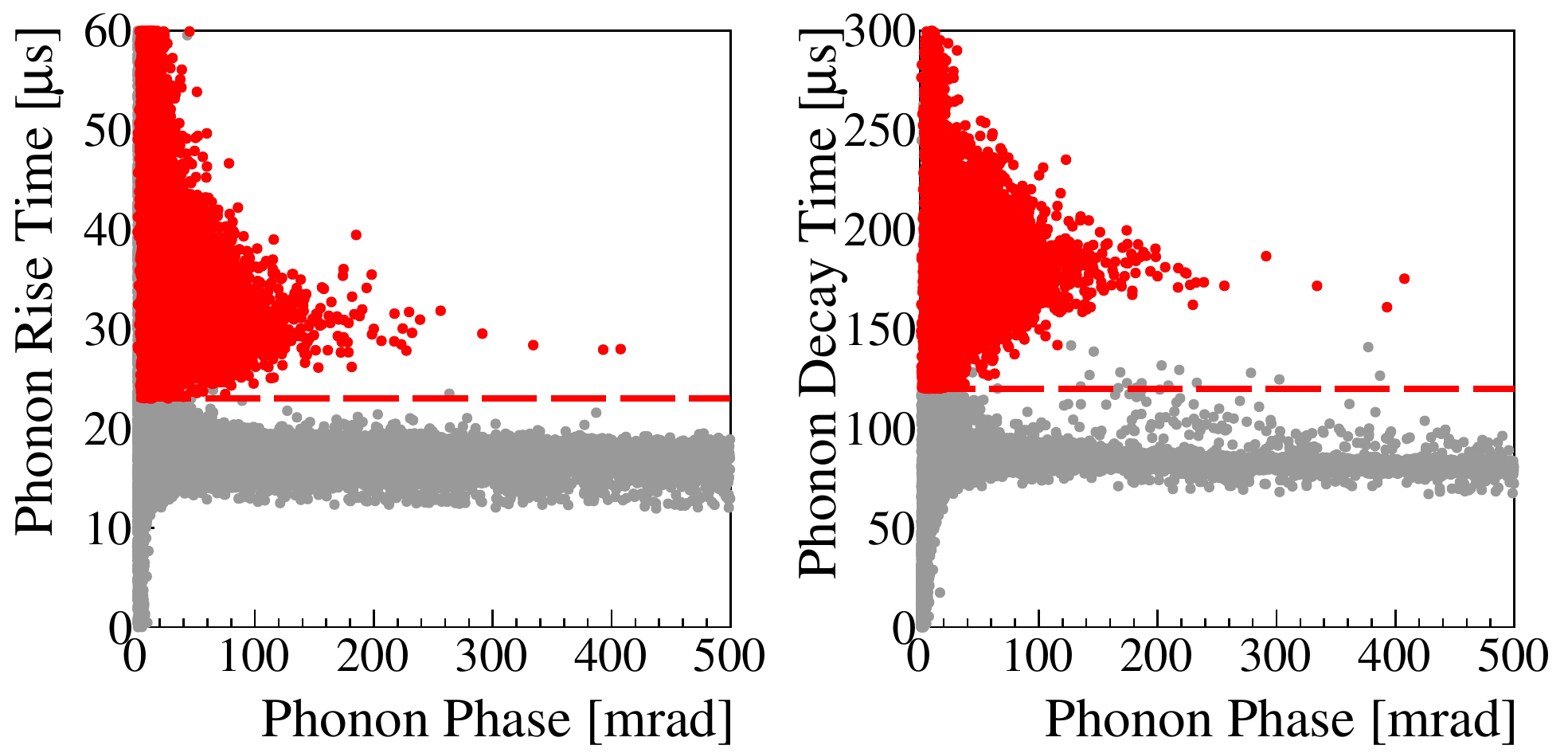}
\caption{Rise time (left) and decay time (right) of the phonon pulses. Two different populations can be easily recognized. The red lines represent the cut values used to select between the two populations: higher values select CR events crossing only the {\LMO} crystal. The events marked in red are selected by requiring that both rise and decay time to be greater than the cut values.}
\label{fig:TimeConstants}
\end{centering}
\end{figure}

As showed in Fig.~\ref{fig:Setup}, the two Si substrates are facing each other. We assembled the detector holder by ensuring that this axis resulted horizontal. Since the majority of cosmic rays (CRs) come from the vertical direction, in such configuration we can maximize the amount of CRs that crossing only the {\LMO} crystal without crossing the Si substrates. Thus, the majority of the interactions will produce both a clean phonon signal in the Si carrier crystal and a clean scintillation light signal in the light detector. Nevertheless, a fraction of CRs will cross both the {\LMO} crystal and one of the two Si substrates (light detector or carrier crystal) depositing energy by ionisation and masking the searched signals. These events must be rejected:
\begin{itemize}
\item if the CR ionizes the Si light detector, the energy released will be of the order of hundreds of keV saturating the KID response. This type of events can be easily tagged. 
\item if the CR crosses the Si carrier crystal glued on to the {\LMO} crystal, the phonon signal will be amplified because of the direct ionization of the Si substrate. As a consequence, the measured phonon signal will be larger with respect to the case in which the CR crosses only the {\LMO} crystal.
\end{itemize}
We expect also a clear difference in the pulse time development of the phonon channel for this last category of events. Indeed, as shown in Fig~\ref{fig:TimeConstants}, we can recognize two populations of pulses triggered by the phonon detector: one with a slower rise and decay time and with a lower maximum energy, and the other faster and with a higher maximum energy. 
\begin{figure}[!t]
\begin{centering}
\includegraphics[width=8.5cm]{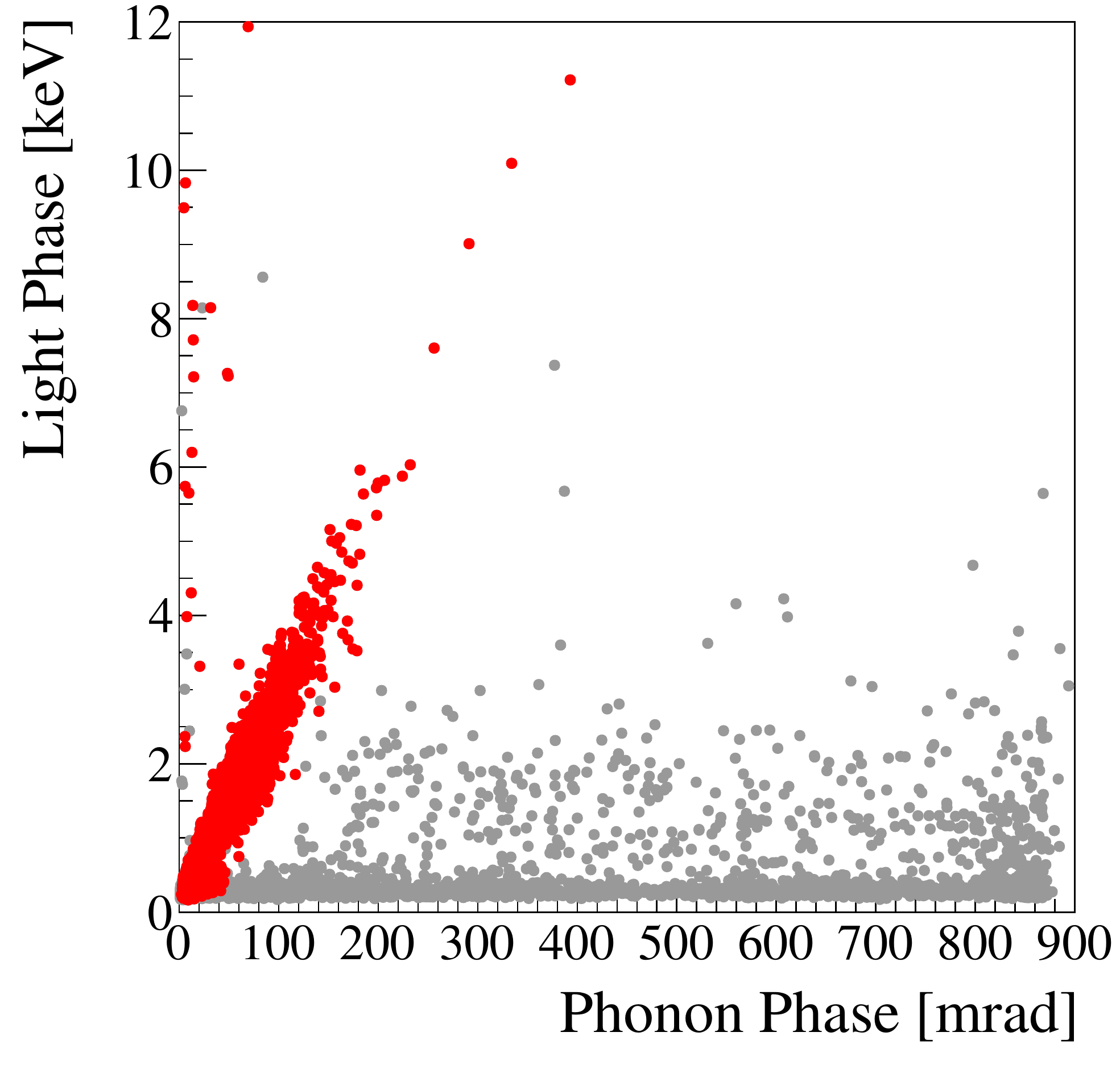}
\caption{Detected light as a function of the phonon signal. The events marked in red are selected by the cuts described in Fig.~\ref{fig:TimeConstants}. As expected, for interactions crossing the {\LMO} crystal the light signal is proportional to the energy release in the crystal. The light detector is energy calibrated exploiting the LED pulses. The phonon channel was not calibrated due to the absence of peaks.}
\label{fig:Scatter}
\end{centering}
\end{figure}
\begin{figure}[!b]
\begin{centering}
\includegraphics[width=8.5cm]{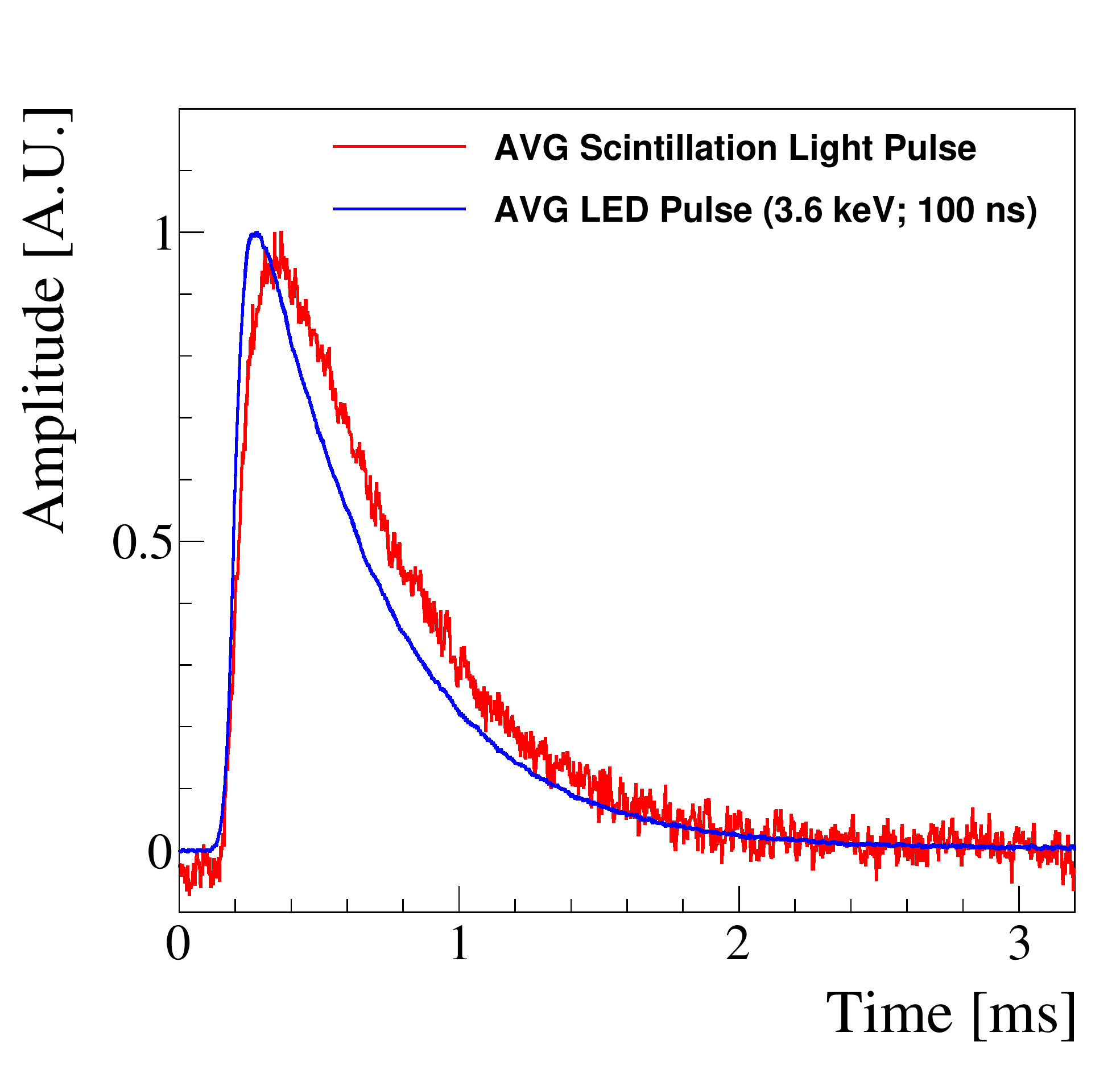}
\caption{Average pulses from LED optical pulses (blue) and scintillation light events (red). The main difference between the two samples is in the pulse rise time. The amplitudes were scaled to one and the time windows were zoomed from the acquired 12~ms to 3.2~ms to simplify the shape comparison.}
\label{fig:AVG}
\end{centering}
\end{figure}
Selecting the pulses with a slower time development through the cut showed in Fig~\ref{fig:TimeConstants}, we obtained the events marked in red in the phonon-light scatter plot of Fig.~\ref{fig:Scatter}. For these events, the light yield measured by the light detector and its trend as function of the energy measured in the phonon channel is compatible with CRs crossing the {\LMO} crystals~\cite{Cardani:2013dia}. The remaining events marked in grey are due to CRs crossing also the Si carrier crystal. Finally, we averaged the waveforms acquired by the light detector in order to reduce random noise fluctuations. We selected only the events marked in red in the scatter plot of Fiq.~\ref{fig:Scatter} with a phonon signal amplitude between 80 and 140~mrad, since they have a light signal with an energy similar to the one deposited with the LED burst. The resulting average pulse is showed in Fig.~\ref{fig:AVG} together with the average pulse made exploiting a 3.6~keV LED burst 100~ns long.

\section{KID response model}
\label{sec:KID response}

The KIDs response to instantaneous energy deposition on Si substrate can be modelled exploiting three time constants~\cite{Martinez:2018ezx}: the time of arrival of phonons ($\tau_{ph}$), i.e. the characteristic time in which the athermal phonons produced in an energy deposition on the substrate arrive to the KID, the time constant of the resonator ($\tau_{ring} = Q/\pi f_0$), and the time in which quasiparticles recombine back into Cooper's pairs ($\tau_{qp}$). The time development of the phase signal is a convolution of these three effects:
\begin{equation}
\label{eq:resp}
\begin{split}
\delta\phi(t) = A\tau_{qp} &\left[ \frac{\tau_{qp}e^{-t/\tau_{qp}}}{(\tau_{qp}-\tau_{ph})(\tau_{qp}-\tau_{ring})}\right.\\
&+\frac{\tau_{ph}e^{-t/\tau_{ph}}}{(\tau_{ph}-\tau_{qp})(\tau_{ph}-\tau_{ring})}\\
&+\left.\frac{\tau_{ring}e^{-t/\tau_{ring}}}{(\tau_{ring}-\tau_{qp})(\tau_{ring}-\tau_{ph})}\right]
\end{split}
\end{equation}
where $A$ is the pulse amplitude. If the energy deposition has a temporal evolution, as is the case of scintillation light, the resulting KID signal is a convolution of Eq.~\ref{eq:resp} with the exponential time development of the scintillation light ($\tau_{scint}$). Thus, the pulses produced by scintillation light are described by:
\begin{equation}
\label{eq:model}
\begin{split}
\delta\phi(t)&=  A\tau_{qp} \left[ \frac{\tau^2_{qp}e^{-t/\tau_{qp}}}{(\tau_{qp}-\tau_{ph})(\tau_{qp}-\tau_{ring})(\tau_{qp}-\tau_{scint})}\right.\\
&-\frac{\tau^2_{ph}e^{-t/\tau_{ph}}}{(\tau_{qp}-\tau_{ph})(\tau_{ph}-\tau_{ring})(\tau_{ph}-\tau_{scint})}\\
&-\frac{\tau^2_{ring}e^{-t/\tau_{ring}}}{(\tau_{qp}-\tau_{ring})(\tau_{ring}-\tau_{ph})(\tau_{ring}-\tau_{scint})}\\
&-\left.\frac{\tau^2_{scint}e^{-t/\tau_{scint}}}{(\tau_{qp}-\tau_{scint})(\tau_{scint}-\tau_{ring})(\tau_{scint}-\tau_{ph})}\right]
\end{split}
\end{equation}
\begin{figure}[!t]
\begin{centering}
\includegraphics[width=8.5cm]{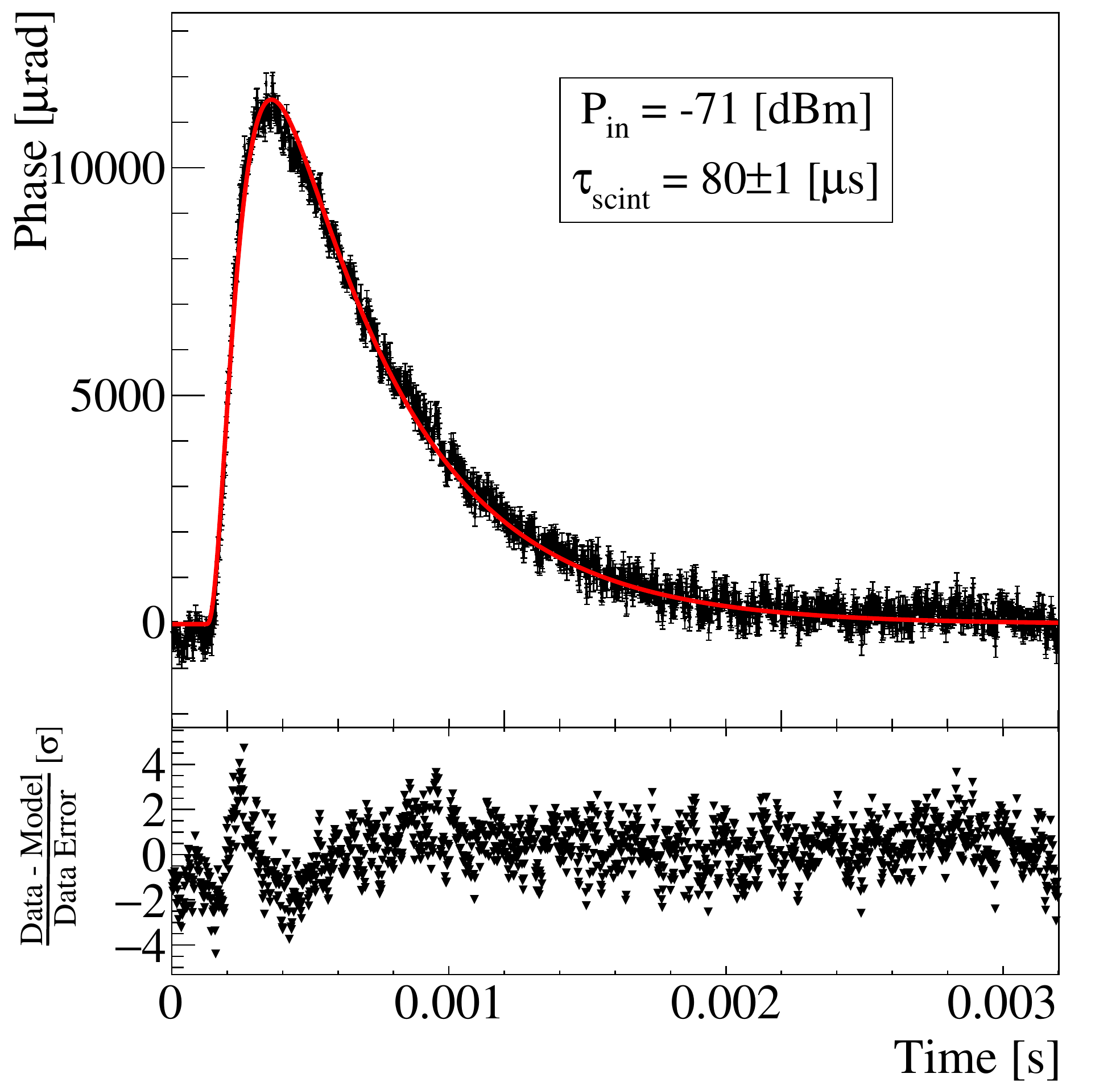}
\caption{Top: average scintillation pulse fitted using Eq.~\ref{eq:model} with $\tau_{qp}$, $\tau_{ph}$ and $\tau_{ring}$ fixed from the fit performed using Eq.~\ref{eq:resp} on the average LED pulse. Bottom: residuals between the data and the model divided by dara error. The microwave input power was -71~dBm.}
\label{fig:AVGfit}
\end{centering}
\end{figure}
\begin{figure}[!b]
\begin{centering}
\includegraphics[width=8.5cm]{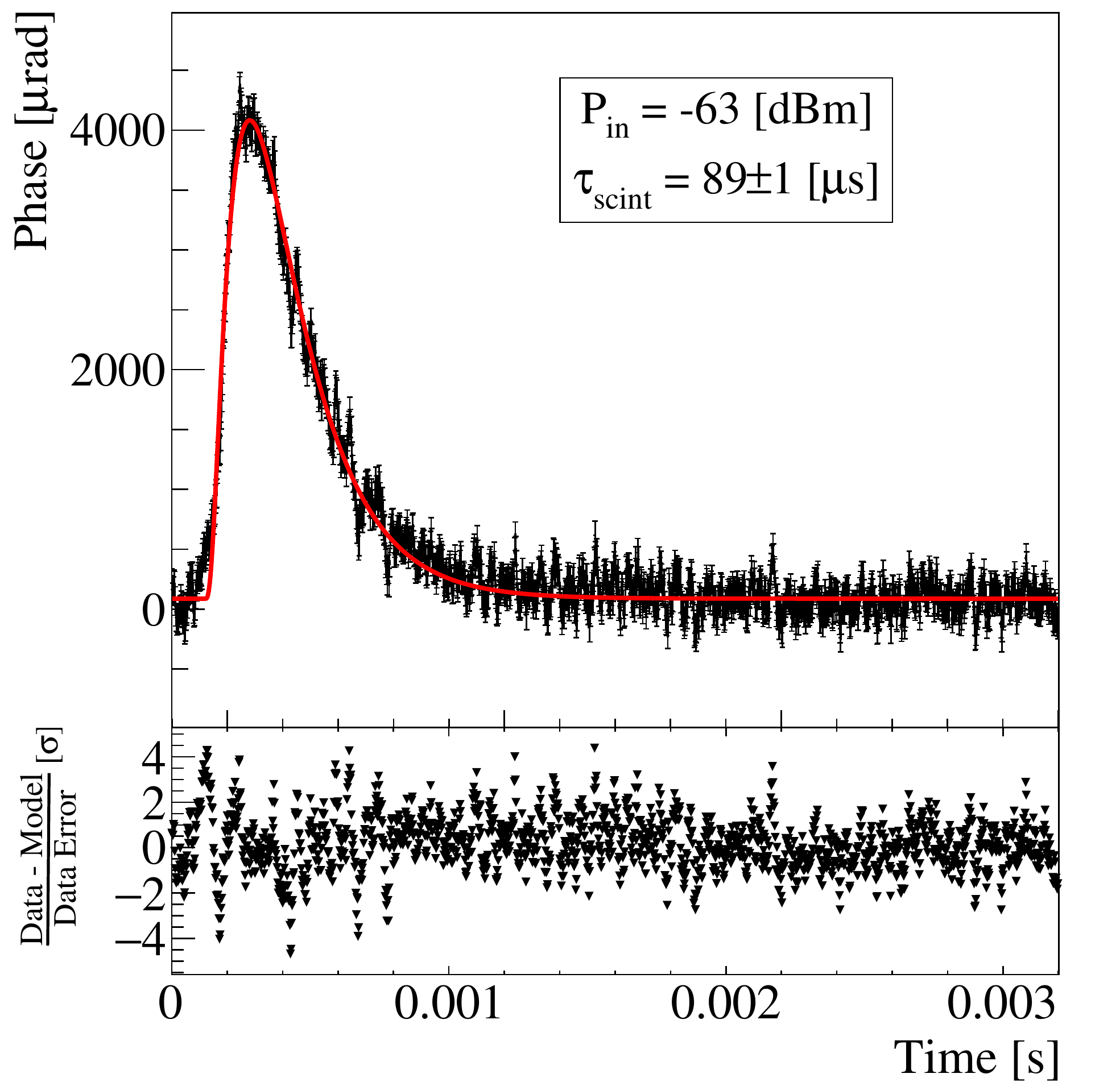}
\caption{Top: average scintillation pulse fitted using Eq.~\ref{eq:model} with $\tau_{qp}$, $\tau_{ph}$ and $\tau_{ring}$ fixed from the fit performed using Eq.~\ref{eq:resp} on the average LED pulse. Bottom: residuals between the data and the model divided by data error. The microwave input power was increased from -71~dBm to -63~dBm in order to decrease $\tau_{qp}$ and than increase the light detector time response as can be easily seen comparing this pulse with the one showed in Fig.~\ref{fig:AVGfit}.}
\label{fig:AVGfit2}
\end{centering}
\end{figure}
According to these models, we use Eq.~\ref{eq:resp} to fit the average LED pulse in order to evaluate $\tau_{ph}$ and $\tau_{qp}$ ($\tau_{ring}$ is fixed to 15~$\mu$s by the resonator parameters) obtaining $\tau_{qp} = 459$~$\mu$s and $\tau_{ph} =38$~$\mu$s.
Then, fixing the three time constants, we fit the average pulse evaluated on the scintillation signal using Eq.~\ref{eq:model} (see Fig.~\ref{fig:AVGfit}) in order to evaluate the remaining $\tau_{scint}$, which results to be $80\pm1$~$\mu$s. In order to understand the systematic error on $\tau_{scint}$ introduced by the procedure used to evaluate this parameter, we increased the input microwave power used to excite the light detector resonator. Indeed, according to Ref.~\cite{Cruciani:2016moq} the microwave power increases the quasi-particle recombination probability decreasing $\tau_{qp}$. In such a way we increase the time response of the light detector decreasing $\tau_{qp}$ by a factor of 2.5. Then, we repeated the analysis procedure on both LED and scintillation average pulses (see Fig.~\ref{fig:AVGfit2}) obtaining $\tau_{scint} = 89\pm1$~$\mu$s. The difference between the two values of $\tau_{scint}$ can be ascribed to approximations in the KID response model~\cite{Martinez:2018ezx} and to a non-perfect pulses alignment in the averaging procedure due to trigger instabilities and noise fluctuations. Nevertheless, we account for these effects as a systematic uncertainty: we average the two time constants and we identified its deviation as the systematic uncertainty of the measurement procedure: $\tau_{scint} = 84.5\pm4.5\rm{(syst)}$. This value is consistent with the one measured at 20~K in Ref.~\cite{CHEN2018225}. Finally we measured $\tau_{scint}$ at 50, 100 and 190~mK with an input power of -71~dBm and did not appreciated variations with respect to the value obtained at 10~mK.
\section{Conclusions}
In this work we applied for the first time the KID technology to macro-absorbers, measuring the phonon and light signals coming from a {\LMO} crystal crossed by cosmic rays.  
We proved the potential of the multiplexed read-out by measuring the two detectors with a single read-out line, and we demonstrated that the carrier technique can be used in this field.

This experiment allowed to measure a parameter of interest for future {\DBD} experiments based on {\LMO}, namely its scintillation time constant at millikelvin temperatures. We accounted for the systematic uncertainty due to the KID response model, obtaining $\tau_{scint} = 84.5\pm4.5\rm{(syst)}\pm1.0\rm{(stat)}$~$\mu$s, in agreement with previous measurements performed with a different technique down to 20~K. 
With a focused R\&D activity on the light detector, it will be possible to decrease the background coming from pile-up events, paving the way to future {\DBD} experiments with an increasing discovery potential.


\begin{acknowledgements}
This work was supported by the European Research Council (FP7/2007-2013) under Contract No. CALDER No. 335359 and by the Italian Ministry of Research under the FIRB Contract No. RBFR1269SL. The authors thank the LUMINEU collaboration for providing a high-quality lithium molybdate crystal, and Th. Redon (CSNSM, Orsay) for the preparation of the sample used in this experiment. The authors also thanks F. Petricca, K. Sch\"affner and N. Ferreiro for the useful discussion about carrier crystal approach and the personnel of INFN Sezione di Roma for the technical support, in particular M. Iannone and A. Girardi.
\end{acknowledgements}

\bibliographystyle{spphys}       

\begin{thebibliography}{10}
\providecommand{\url}[1]{{#1}}
\providecommand{\urlprefix}{URL }
\expandafter\ifx\csname urlstyle\endcsname\relax
  \providecommand{\doi}[1]{DOI \discretionary{}{}{}#1}\else
  \providecommand{\doi}{DOI \discretionary{}{}{}\begingroup
  \urlstyle{rm}\Url}\fi


\bibitem{Furry}
W.~H.~Furry,
  Phys.Rev.Lett. \textbf{56}, 1184 (1936).

\bibitem{Feruglio:2002af} 
F.~Feruglio, A.~Strumia and F.~Vissani, Nucl.\ Phys.\ B {\bf 637}, 345 (2002)  Addendum: [Nucl.\ Phys.\ B {\bf 659}, 359 (2003)].
\newblock \doi{10.1016/S0550-3213(02)00345-0,10.1016/S0550-3213(03)00228-1}\

\bibitem{Strumia:2005tc} 
A.~Strumia and F.~Vissani, Nucl.\ Phys.\ B {\bf 726}, 294 (2005).
\newblock \doi{10.1016/j.nuclphysb.2005.07.031}

\bibitem{Barabash:2019nnr} 
  A.~S.~Barabash,
  \newblock \doi{1907.06887}

\bibitem{Wang:2015taa}
G.~Wang, et~al., arXiv:1504.03612  (2015)



\bibitem{Artusa:2014lgv}
D.R. Artusa, et~al., Adv. High Energy Phys. \textbf{2015}, 879871 (2015).
\newblock \doi{10.1155/2015/879871}







\bibitem{Alduino:2017qet} 
C.~Alduino {\it et al.} [CUORE Collaboration], Eur.\ Phys.\ J.\ C {\bf 77} no.8,  543 (2017).\newblock \doi{10.1140/epjc/s10052-017-5080-6}.

\bibitem{Azzolini:2018tum}
O.~Azzolini et~al. [CUPID Collaboration], Eur. Phys. J. C {\bf 78} 428 (2018). 
\newblock \doi{10.1140/epjc/s10052-018-5896-8}

\bibitem{Armengaud:2017hit}
E.~Armengaud et~al., Eur.\ Phys.\ J.\ C {\bf 77}  no.11,  785 (2017).
\newblock \doi{10.1140/epjc/s10052-017-5343-2}

\bibitem{Alenkov:2019jis} 
  V.~Alenkov {\it et al.},
  \newblock \doi{arXiv:1903.09483}.

\bibitem{Barabash:2011}
A. Barabash, Phys. Rev. C 81, 035501, (2011).
\newblock \doi{10.1103/PhysRevC.81.035501}

\bibitem{Azzolini:2018yye} O.~Azzolini {\it et al.}, Eur.\ Phys.\ J.\ C {\bf 78}  no.9, 734 (2018). \newblock \doi{10.1140/epjc/s10052-018-6202-5}

\bibitem{Chernyak1}	
D.~M.~Chernyak et al., Eur. Phys. J. C \textbf{74}, 2913 (2014)
\newblock \doi{10.1140/epjc/s10052-014-2913-4}


\bibitem{Chernyak2}	
D.~M.~Chernyak et al., Eur. Phys. J. C \textbf{77}, 3 (2017)
\newblock \doi{10.1140/epjc/s10052-016-4565-z}

\bibitem{Barucci:2019ghi} 
  M.~Barucci et al., Nucl.\ Instrum.\ Meth.\ A {\bf 935}, 150 (2019)
\newblock \doi{10.1016/j.nima.2019.05.019}

\bibitem{CHEN2018225}
P.~Chen et~al., Materials Letters \textbf{215} 225 - 228 (2018).
\newblock \doi{10.1016/j.matlet.2017.12.113}

\bibitem{bay}
P.~K.~Day, H.~G.~LeDuc, B.~A.~Mazin, A.~Vayonakis, and J.~Zmuidzinas. Nature, \textbf{425}, 817-821 (2013).
\newblock \doi{10.1038/nature02037}

\bibitem{vanRantwijk:2015sta} 
  J.~van Rantwijk, M.~Grim, D.~van Loon, S.~Yates, A.~Baryshev and J.~Baselmans,
  IEEE Trans.\ Microwave Theor.\ Tech.\  {\bf 64}, 1876 (2016).
 \newblock \doi{10.1109/TMTT.2016.2544303}
  
\bibitem{Mazin:2010pz}
  B.~A.~Mazin, K.~O'Brien, S.~McHugh, B.~Bumble, D.~Moore, S.~Golwala and J.~Zmuidzinas,
  Proc.\ SPIE Int.\ Soc.\ Opt.\ Eng.\  {\bf 7735} (2010) 773518
  \newblock \doi{10.1117/12.856440}

\bibitem{Swenson:2010}
  L.~J.~Swenson, et~al.,
 J. Appl. Phys. 96 (2010) 263511.
\newblock \doi{10.1063/1.3459142}

\bibitem{Moore:2012}
  D.~C.~Moore, et~al.,
 J. Appl. Phys. 100 (2012) 232601.
\newblock \doi{10.1063/1.4726279}

\bibitem{Battistelli:2015vha}
E.~Battistelli, et~al.,
 Eur. Phys. J. C 75, 8 (2015) 353.  
\newblock \doi{10.1140/epjc/s10052-015-3575-6}

\bibitem{Bellini:2016lgg} 
  L.~Cardani, et~al.,  Appl. Phys. Lett.  {\bf 110}, no. 3, 033504 (2017).
  \newblock \doi{10.1063/1.4974082}

\bibitem{cruciani2018}
L.~Cardani et~al., 
  Supercond. Sci. Technol. 31 (2018) 075002.
 \newblock \doi{10.1088/1361-6668/aac1d4}


\bibitem{cresst}	
R~Strauss, et~al., Eur. Phys. J. C \textbf{75}, 352 (2015) . 
\newblock\doi{10.1140/epjc/s10052-015-3572-9}

\bibitem{tes}	
J.~Rothe, et~al., J. Low Temp. Phys. (2018). 
\newblock\doi{10.1007/s10909-018-1944-x}

\bibitem{ampli}
http://radiometer.caltech.edu/datasheets/amplifiers/CITLF4.pdf

\bibitem{Bourrion:2011gi} 
  O.~Bourrion, et al.,
  JINST {\bf 6}, P06012 (2011).
  \newblock\doi{10.1088/1748-0221/6/06/P06012}

\bibitem{Gatti:1986cw}
E.~Gatti, P.F. Manfredi, Riv. Nuovo Cimento \textbf{9}, 1 (1986)

\bibitem{Radeka:1966}
V.~Radeka, N.~Karlovac, Nucl. Instrum. Meth. \textbf{52}, 86 (1967)
   
   \bibitem{Cardani:2013dia} 
  L.~Cardani, et~al.,
  JINST {\bf 8}, P10002 (2013)
  \newblock\doi{10.1088/1748-0221/8/10/P10002}
   
\bibitem{Martinez:2018ezx} 
  M.~Martinez, L.~Cardani, N.~Casali, A.~Cruciani, G.~Pettinari and M.~Vignati,
  Phys.\ Rev.\ Applied.\  {\bf 11}, 064025 (2019).
 \newblock \doi{10.1103/PhysRevApplied.11.064025}
 
\bibitem{Cruciani:2016moq} 
  A.~Cruciani, et~al.,
  J.\ Low.\ Temp.\ Phys.\  {\bf 184}, no. 3-4, 859 (2016).
  \newblock \doi{10.1007/s10909-016-1574-0}
 
  
\end{thebibliography}

\end{document}